# Current-induced magnetization reversal in a (Ga,Mn)As-based magnetic tunnel junction

Rai Moriya[1], Kohei Hamaya[2], Akira Oiwa[3], and Hiro Munekata[1*]

*E-mail address: hiro@isl.titech.ac.jp

[1]Imaging Science and Engineering Laboratory, Tokyo Institute of Technology

4259-G2-13 Nagatsuta-cho, Midori-ku, Yokohama 226-8502, Japan.

[2]Department of Innovative and Engineered Materials, Tokyo Institute of Technology

4259-G1-29 Nagatsuta-cho, Midori-ku, Yokohama 226-8502, Japan.

[3]PRESTO, Japan Science and Technology Agency, 4-1-8 Honcho, Kawaguchi 332-0012, Japan.

**Abstract**

We report current-induced magnetization reversal in a ferromagnetic semiconductor-based magnetic tunnel junction (Ga,Mn)As/AlAs/(Ga,Mn)As prepared by molecular beam epitaxy on a $p$-GaAs(001) substrate. A change in magneto-resistance that is asymmetric with respect to the current direction is found with the excitation current of $10^6$ A/cm$^2$. Contributions of both unpolarized and spin-polarized components are examined, and we conclude that the partial magnetization reversal occurs in the (Ga,Mn)As layer of smaller magnetization with the spin-polarized tunneling current of $10^5$ A/cm$^2$.

When an electrical current of spin-polarized carriers enters a ferromagnetic film, the transfer of angular momentum occurs between the carriers and the local magnetic moments in the film. This phenomenon, which is called spin transfer,[1] has served a theoretical base for the proposal of the magnetization reversal by the spin-polarized current that flows between the two ferromagnetic films.[2,3] The electrical power needed to reverse the magnetization is expected to decrease with reducing the size of magnetic binary bits, which is quite favorable for the development of spintronics devices such as magnetic random access memory. Experiments using metallic nano-structures showed that magnetization could be reversed with the current density of $10^{7-8}$ A/cm$^2$,[4,5,6] which, however, is still high to be used in the integrated circuits.

In this letter, we report the current-induced magnetization reversal (CIMR) in a III-V-based ferromagnetic semiconductor (Ga,Mn)As for which reduced current density is expected because of its small magnetization value. Experimental evidence of the CIMR is found by studying a change in magneto-resistance (MR) after switching on the excitation current between two magnetically-different (Ga,Mn)As layers through an AlAs tunnel barrier. The change in MR is asymmetric with respect to the direction of the current flow of $10^6$ A/cm$^2$. The contributions of both unpolarized and spin-polarized components are carefully studied, through which we have concluded that one-fourth of the magnetization in the thinner (Ga,Mn)As layer is reversed by the spin-polarized tunneling current of $10^5$ A/cm$^2$.

Epitaxial $p$-(Ga,Mn)As/AlAs/$p$-(Ga,Mn)As magnetic tunnel junctions (MTJs) were prepared by molecular beam epitaxy on $p$-type GaAs(001) substrates. The MTJ was used because this structure has been proven to exhibit a large MR for (Ga,Mn)As based nano-structures,[7] which is advantageous to detect a change in the relative magnetization configuration between the two ferromagnetic layers. On the other hand, a large bias is anticipated in order to switch on a sufficient amount of spin polarized current across the junction. As shown schematically in Fig. 1(a), the actual MTJ structure consisted of, from the top surface, a 30-nm Ga$_{0.96}$Mn$_{0.04}$As, a 1-nm GaAs, a 1.8-nm AlAs, a 1-nm GaAs, a 8-nm

Ga$_{0.95}$Mn$_{0.05}$As, a 300-nm GaAs:Be, and a $p^+$-GaAs:Zn (100) substrate. The Curie temperatures for the top and the bottom (Ga,Mn)As layers were $T_C \sim 50$ K and 30 K, respectively. The thicknesses and Mn contents of the top and bottom (Ga,Mn)As layers were determined so as to introduce the difference in the magnetization and the coercive force between the two layers. The 1-nm thick GaAs spacers were introduced to achieve flat hetero-interfaces.[7] The barrier height for holes at the AlAs/(Ga,Mn)As interface is inferred to be 0.65 eV for holes.[8,9]

The starting epilayer sample was processed into a circular mesa structure of the diameter $d = 1$ μm and the height $h = 200$ nm. A circular mesa was used since the shape-induced magnetic anisotropy would not become dominant in the (Ga,Mn)As-based structure of 1-μm size.[10] As shown in Fig. 1(b), a 300-nm thick, Au/Ti electrode of the diameter $d = 1$ μm was fabricated on top of the sample by using standard electron beam lithography and a lift-off process. This was followed by the mesa fabrication with chemical wet etching, and the device isolation with a 1-μm thick silicate polymer glass (SPG; OCD T-7, Tokyo Ohka). Finally, a contact hole was fabricated by selectively removing the polymer layer on top of the Au/Ti electrode with the focused ion beam (FIB) technique, and the second Au/Ti metal pad was placed through the hole. The acceleration voltage and Ga$^+$ ion current were 30 keV and 0.6 nA, respectively.

Figure 2 shows a magnetization hysteresis curve (a M-H curve) of the starting epilayer sample and a magneto-resistance hysteresis curve (a MR curve) of the 1-μm diameter device at 10 K. The MR curve was measured at the constant bias voltage of 5 mV at which the current density was around $10^2$ A/cm$^2$. The M-H curve exhibits a typical two-step behavior attributed to a change between parallel and anti-parallel magnetization configuration as a function of an external magnetic field. Being consistent with the M-H curve, the peaks appearing at $H \sim \pm 170$ Oe in the MR curve are the manifestation of the anti-parallel type magnetic configuration. The MR values at the peaks are about 5 %, being smaller than the reported value of 30 % for the similar structure.[7] As discussed in the next paragraph, we attribute the small MR value to the contribution of a leak current.

As shown in the inset, the current-voltage (*I-V*) curve of the *d* = 1 μm device exhibits symmetric, non-linear feature, being characteristic of the single barrier tunneling. However, the large current density, being quite distinct compared with the *I-V* curve (thick solid line in the inset) of the *d* = 140 μm device,[11] indicates the presence of leak current paths in the *d* = 1 μm device. The MR value is also larger (MR = 30 %) in the *d* = 140 μm device. The analysis with the single-barrier-tunneling model[12] for these two samples yields the barrier height and the thickness of $\Delta E_V$ = 0.45 eV and *t* = 1.3 nm, and $\Delta E_V$ = 0.54 eV and *t* = 2.02 nm,[13] for *d* = 1 μm and 140 μm, respectively, which also supports the presence of a leak current in the *d* = 1 μm device. We infer that, nearly at the end of the FIB process when the uncovered Au/Ti electrode was bombarded by the $Ga^+$ ions, some moderate leakage paths were generated in the MTJ region. Deposition of an additional protection layer on top of the Au/Ti electrode would be one of the reasonable solutions to impede this problem.

The influence of the excitation current on magnetization reversal was studied experimentally by the following sequence. Firstly, the in-plane magnetic field of *H* = 500 Oe was applied to realize the parallel magnetization configuration. This is represented by the point A of the MR curve shown by dotted lines in Fig. 3. The field was then swept down to the magnetic field (*H* = − 200 Oe) at which the anti-parallel magnetization configuration was realized. This manifested itself as the highest MR value of the point B in Fig. 3. This was followed by switching on either positive (top-to-bottom) or negative (bottom-to-top) excitation current for 1 sec at *H* = − 200 Oe. The reduction in the MR value is expected if the anti-parallel configuration is affected by the excitation current. Finally, the MR curve was measured again at the bias voltage of 5 mV with increasing a magnetic field toward the point A. This step, being equivalent to a minor loop, allows us to study the overall magnetic configuration after the excitation.

We show the MR data obtained after the current excitation with ±$I_{exc}$ ($I_{exc}$ = 1 × $10^6$ A/$cm^2$) in Fig. 3. Here, circles and squares represent the data with the excitation of +$I_{exc}$ and $-I_{exc}$, respectively. The bias voltage of 0.8 V was needed to let this amount of current flow through the device. The MR curve (curve c) obtained after heat-cycling the device between 10 and 20 K without $I_{exc}$ is also shown in the figure for comparison. The data obtained by the $-I_{exc}$ excitation coincide well with the MR curve c. This fact indicates that thermally assisted magnetization reversal is the dominant mechanism for the $-I_{exc}$ excitation. To be striking, the amount of reduction by the $+I_{exc}$ excitation (point D) is significantly larger than that induced by the $-I_{exc}$ excitation (point C). This behavior clearly indicates the presence of some extra mechanism in addition with the thermal assisted magnetization reversal. The contribution of a current-induced magnetic field could be negligible, since it was 28 Oe on the side of the 1-μm mesa with the current flow of 7.5 mA ($I = 1 \times 10^6$ A/cm$^2$). This field is significantly smaller than the coercive forces of the constituent (Ga,Mn)As layers and the external magnetic field of $H = -200$ Oe. These analyses lead us to conclude that the only remaining mechanism that is responsible for the extra magnetization reversal is the spin transfer mechanism induced by the spin-polarized current. The value of the spin-polarized current is estimated to be $1 \times 10^5$ A/cm$^2$ referring to the *I-V* curve (the broken line in the inset of Fig. 2) calculated by using tunneling parameters extracted from the leak-free $d = 140$ μm device.

In view of magnetization, the difference in resistance between the points C and D indicates the relative amount of change from the anti-parallel to the parallel configuration to be about 25 %. In the light of spin transfer model,[2] we infer that it was the magnetization of the thinner (8-nm) Ga$_{0.95}$Mn$_{0.05}$As layer that was partially reversed to form the multi-domain structure by the spin-polarized tunneling current of $1 \times 10^5$ A/cm$^2$. As for the MR minor loops (plots a and b) obtained after the current excitation, they are accompanied by the MR peaks whose magnitudes and field positions differ from those before the excitation. These changes suggest that the magnetic structure of the thick (Ga,Mn)As layer was also affected by the excitation current, specifically, by the current-induced heating. Nevertheless, we were able to reset the device at the point A and reproduce the behavior shown in Fig. 3 for many times.

Reflecting a small magnetization value, the critical current for magnetization reversal in a 8-nm $Ga_{0.95}Mn_{0.05}As$ is estimated to be as small as middle $10^4$ A/cm$^2$ on the basis of the spin transfer model,[2] assuming the spin polarization of $P = 0.85$[14] and the Gilbert damping factor of $\alpha = 0.02$.[15] This estimation leads us to believe that, in parallel with studying the suppression of an unpolarized leak current, there is a room to further reduce the critical current in the (Ga,Mn)As-based MTJ. For the MTJs, a relatively high bias voltage is needed to switch on a sufficient amount of current between the two magnetic layers. On the other hand, the spin injection efficiency $g$ is supposed to reduce with increasing the bias voltage, since injected hot carriers quickly loose their spin information through magnon excitation or impurity-assisted tunneling.[16] In fact, in the present case, the $g$ value is estimated to be as low as $g = 0.06$ for the tunneling current of $1 \times 10^5$ A/cm$^2$ at the bias voltage of 0.8 V. This problem can be solved by either reducing the barrier height or the barrier thickness. Reduction of the barrier height would be the realistic solution in view of the heterostructure fabrication. Realization of the device consisting of a single domain (Ga,Mn)As is another important device optimization, by which the current that is used for the domain-wall switching in the multi-domain structures can be eliminated. The amplified effect found recently in the magnetization rotation by the optical spin injection[17,18] may also be utilized to further reduce the current density below $10^4$ A/cm$^2$, if one finds the way to realize the similar effect by the electrical excitation.

In conclusion, we have demonstrated the current-induced magnetization reversal in a magnetic tunnel junction composed of a III-V-based ferromagnetic semiconductor (Ga,Mn)As. Effects associated with both unpolarized and spin-polarized currents have been discussed, and we have concluded that one-fourth of the magnetization in the thinner (Ga,Mn)As layer was reversed by the spin-polarized tunneling current of $10^5$ A/cm$^2$.

Authors gratefully acknowledge Prof. Y. Kitamoto for the collaboration in (Ga,Mn)As nano-structure fabrication. This work is supported in part by the Grant-in-Aid for Scientific Research in Priority Areas "Semiconductor Nano-spintronics" of The Ministry of Education,

Culture, Sports, Science and Technology (MEXT), Japan. R. M. acknowledges partial support from the 21st Century COE Program "Photonics Nanodevice Integration Engineering" of MEXT, Japan.

**Figure Captions**

Figure 1 : Schematic illustrations of cross sectional view of (a) (Ga,Mn)As-based epitaxial MTJ and (b) MTJ device used for the experiments. Mesa structure was fabricated by the wet etching. Vertical arrows in the device represent the direction of excitation current $\pm I_{exc}$.

Figure 2 : A magnetization hysteresis curve of the starting epilayer sample and a magneto-resistance hysteresis curve of the 1-μm diameter device at 10 K. Magnetic fields were applied in the sample plane, parallel to the [100] direction. Inset shows current-voltage (*I-V*) characteristics at 10 K for 1-μm (thin solid line) and 140-μm (thick solid line) diameter devices. The broken line represents the *I-V* curve calculated based on $\Delta E_V = 0.54$ eV and $t = 2.02$ nm.

Figure 3 : A TMR major loop (dotted line) of a 1-μm diameter device at 10 K before the electrical excitation, together with three different minor loops measured after the excitation with a current of $+I_{exc}$ (plot a), $-I_{exc}$ (plot b), and after heat cycling between 10 and 20 K. Arrows A ~ D represent four different magnetic configurations; (A) parallel magnetization configuration, (B) anti-parallel configuration, (C) the configuration after the excitation with $-I_{exc}$, and (D) the configuration after the excitation with $+I_{exc}$. Magnetic fields were applied in the sample plane, parallel to the [100] direction.

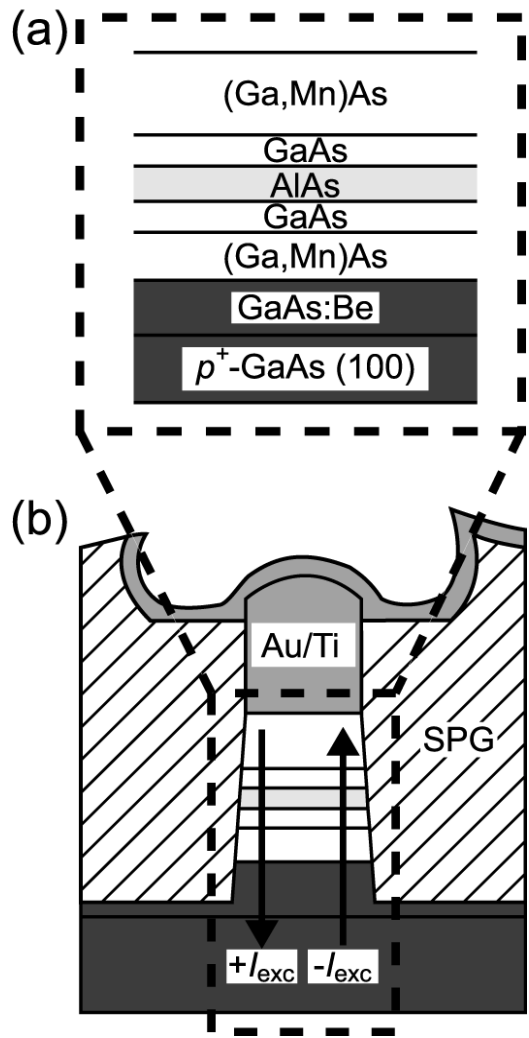

4.5 cm

R. Moriya *et al.*
Fig.1

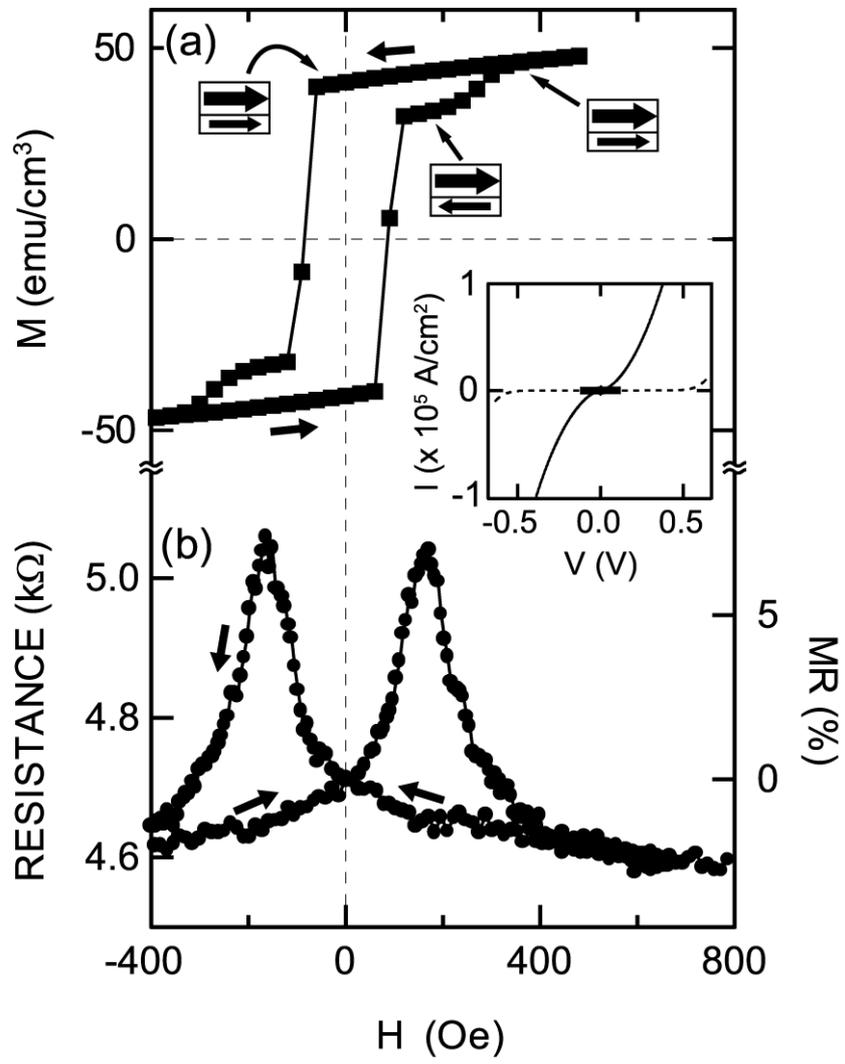

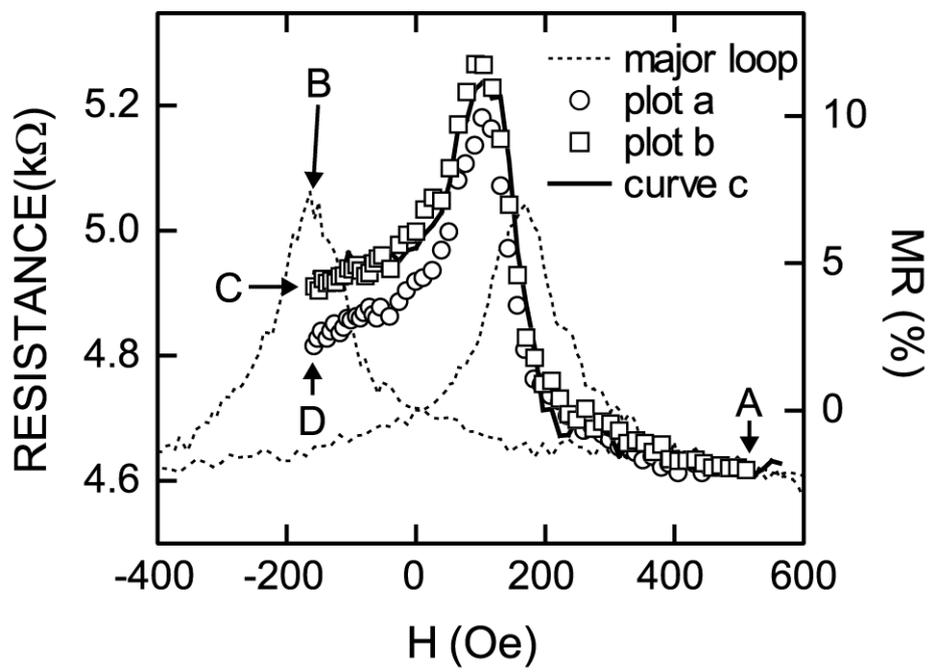

8.2 cm

R. Moriya *et al.*
Fig. 3